\documentclass[aps,twocolumn,nofootinbib,notitlepage,superscriptaddress]{revtex4}
\usepackage{graphicx}
\usepackage{dcolumn}
\usepackage{bm}
\usepackage{amsmath,amssymb}
\usepackage{float}
\usepackage{multirow}
\usepackage{slashed}
\usepackage{xcolor}
\usepackage{physics}
\usepackage{multirow}
\usepackage[colorlinks=true, pdfstartview=FitV, bookmarks=true, bookmarksnumbered=true, breaklinks]{hyperref}
\usepackage{mathtools,braket}
\usepackage{soul}
\usepackage{lipsum}  
\usepackage{color}
\definecolor{blue}{rgb}{0.0, 0.0, 1.0}
\definecolor{red}{rgb}{1.0, 0.0, 0.0}
\definecolor{royalblue}{rgb}{0.0, 0.14, 0.4}
\hypersetup{linkcolor=royalblue, citecolor=blue, urlcolor=royalblue}

\usepackage{hyperref}
\hypersetup{colorlinks=true,citecolor=blue,linkcolor=blue,urlcolor=blue}

\usepackage[mathlines]{lineno}
\usepackage{ulem}

\usepackage{tikz,xcolor,hyperref}
\definecolor{lime}{HTML}{A6CE39}
\DeclareRobustCommand{\orcidicon}{%
	\begin{tikzpicture}
	\draw[lime, fill=lime] (0,0) 
	circle [radius=0.16] 
	node[white] {{\fontfamily{qag}\selectfont \tiny ID}};
	\draw[white, fill=white] (-0.0625,0.095) 
	circle [radius=0.007];
	\end{tikzpicture}
	\hspace{-2mm}
}

\foreach \x in {A, ..., Z}{%
	\expandafter\xdef\csname orcid\x\endcsname{\noexpand\href{https://orcid.org/\csname orcidauthor\x\endcsname}{\noexpand\orcidicon}}
}
\begin{document}
%======================================================
\title{Nuclear medium meson structures from the Schwinger proper-time Nambu--Jona-Lasinio model}
%======================================================

%======================================================
\author{Geoffry Gifari\orcidD}
\email{geoffrygifari.work@gmail.com}
\affiliation{Departemen Fisika, FMIPA, Universitas Indonesia, Depok 16424, Indonesia}
%------------------------------------------------------
\author{Parada~T.~P.~Hutauruk\orcidA}
\email{phutauruk@pknu.ac.kr}
\affiliation{Department of Physics, Pukyong National University (PKNU), Busan 48513, Korea}
%------------------------------------------------------
%------------------------------------------------------
\author{Terry Mart\orcidC}
\email{terry.mart@sci.ui.ac.id}
\affiliation{Departemen Fisika, FMIPA, Universitas Indonesia, Depok 16424, Indonesia}
%======================================================
\date{\today}

%======================================================
\begin{abstract}
In this paper, we report the results of our study on the nuclear medium modifications of the meson electromagnetic form factors in the framework of the Nambu--Jona-Lasinio (NJL) model with the help of the Schwinger proper-time regularization scheme to tame the loop divergence and simulate the effect of QCD confinement. In our current approach, the meson structure and nuclear medium are constructed in the same NJL model at the quark level. We examine the free-space and in-medium charge radii of kaon and pion, the spacelike elastic electromagnetic form factors of kaon and pion, and their quark-sector form factors, which reflect their internal structures. By comparing our result to experimental data, we found that the free-space elastic electromagnetic form factors of the mesons are consistent with the data, while the in-medium elastic electromagnetic form factors of the mesons are found to decrease as the nuclear matter density increases, leading to a rise of meson charge radius, which is consistent with the prediction of other theoretical calculations. We also predict the axial nucleon coupling constant $g_A$ in nuclear medium computed via the Goldberger-Treiman relation (GTR), which is crucial for searching the neutrinoless double beta decay ($0\nu \beta \beta$).
\end{abstract}
%======================================================
\maketitle

%======================================================
\section{Introduction} \label{sec:intro}
%======================================================
Kaons and pions play an important role in the description of the low-energy dynamics and properties of the strong interactions nonperturbative QCD~\cite{Ioffe:2005ym} such as color QCD confinement and dynamical chiral symmetry breaking (D$\chi$SB), which are related to the so-called hadron mass generation or emergent hadron mass (EHM)~\cite{Accardi:2023chb}. In the nuclear medium, it is expected that the partial restoration of the chiral symmetry breaking occurs at higher nuclear matter density. However, the question of how the restoration mechanism happens in the nuclear medium and how it affects the structure of hadron remains unsolved and is still poorly understood. It is widely known that not only are the weak properties of hadrons modified in a nuclear medium but also the structure of hadrons is expected to change, as observed for the first time by the Stanford Linear Accelerator Center (SLAC) experiment, which is then so-called the European Muon Collaboration (EMC) effect~\cite{EuropeanMuon:1983wih,Geesaman:1995yd}. 

In the past many studies on kaon and pion properties in nuclear medium have been made within the various theoretical models, such as the hybrid light-front--quark-meson coupling (LF-QMC) model~\cite{deMelo:2014gea}, the QCD sum rules~\cite{Er:2022cxx}, the hybrid light-front constituent quark--quark-meson coupling (LFCQM-QMC) model~\cite{Arifi:2023jfe}, and the hybrid Nambu--Jona-Lasino--quark meson coupling (NJL-QMC) model~\cite{Hutauruk:2019was,Hutauruk:2018qku}. In the literature, most of those studies and attempts have made use of the hybrid model, meaning in their calculation they combined two different models in calculating the nuclear matter and the structure of hadron. For instance, the authors of Ref.~\cite{deMelo:2014gea} studied the pion structure using the LF-QMC model, where the LF model is used to calculate the pion structure in a relativistic manner, on one hand, and the QMC model is used to calculate the quark mass properties in the nuclear medium, on the other hand. It is worth mentioning that in the QMC model, the light quark is coupled to the strong scalar and vector-meson mean fields to yield the medium quark masses. However, the dynamical spontaneous chiral symmetry breaking that dynamically generates the constituent quark mass is not clearly described in both LF and QMC models. The authors of Ref.~\cite{Whittenbury:2016sma} have tried to capture this quark condensate in the QMC model through the NJL model, where dynamical symmetry breaking is well explained in the model. In the present study, we consistently calculate the elastic electromagnetic form factors of the meson and nuclear matter in the NJL model, where, in the NJL model, dynamical spontaneous chiral symmetry breaking and its partial restoration in the nuclear medium are captured in the chiral quark condensate as the quantity of the order parameter.

In this paper, we evaluate the kaon and pion properties and their medium modifications of the electromagnetic form factors using the NJL model with the help of the Schwinger proper-time regularization scheme. In the calculation, in addition to the meson structure, the nuclear medium is also computed using the NJL model. The NJL model is an uncomplicated model and a very useful tool. The NJL model has been widely and successfully used to describe a number of physics phenomena such as kaon and pion parton distribution functions in free space and medium~\cite{Hutauruk:2016sug,Hutauruk:2019ipp}, gluon distribution functions for the kaon and pion~\cite{Hutauruk:2021kej,Hutauruk:2023ccw}, gluon EMC effects in nuclear matter~\cite{Wang:2021elw,Hutauruk:2022zju}, color superconducting in neutron star~\cite{Tanimoto:2019tsl,Bentz:2001vc,Hutauruk:2023mjj}, charge symmetry breaking effect in pion and kaon structures~\cite{Hutauruk:2018zfk}, transverse momentum dependent (TMD) quark distribution function~\cite{Ninomiya:2017ggn}, transverse momentum dependent fragmentation functions~\cite{Bentz:2016rav}, pion cloud effect on the nucleon form factors~\cite{Cloet:2014rja}. Very recently, the NJL model was used to study the charge symmetry-breaking effect on the parton distribution functions~\cite{Hutauruk:2022sbm} and pseudoscalar properties at finite nuclear matter density and temperature~\cite{Hutauruk:2021dgv}. 

In this work, we found that the in-medium modifications of the electromagnetic form factors for kaons in nuclear matter decrease as the nuclear matter density increases, as expected. On the contrary, we found that the charge radii of the pion and kaon increase as the nuclear matter density increases, which is consistent with other predictions~\cite{Er:2022cxx,Hutauruk:2019was,Hutauruk:2018qku}. At nuclear matter density $\rho_B = \rho_0$, it is estimated that the enhancements of the charge radii for the pion, positively charged, and neutral kaon are approximately 10.33\%, 8.87\%, and 21.69\%, respectively, in comparison to those in free space. Similarly, for elastic electromagnetic form factors, the decay constant, and mass of the pion decrease as the nuclear density increases. On the contrary, the kaon decay constant and mass in the nuclear medium are found to increase as the nuclear matter density increases.

In the experiment sector, the results of our current study provide very useful information for future experiments that plan to measure the properties of the quark and meson in the higher nuclear matter density (nuclear chemical potential) at Compressed Baryonic Matter Experiment (CBM) at Antiproton and Ion Research (FAIR) in Germany, the Nuclotron Ion Collider Facility (NICA) at the JINR in Russia, and at (Japan Proton Accelerator Research Complex (J-PARC) in Japan. Furthermore, the results for weak axial-vector coupling $g_A$ in the nuclear medium are very important for searching the neutrinoless double beta decay ($0\nu\beta \beta$) and weak interaction nuclear processes~\cite{Suhonen:1998ck,Cirigliano:2022rmf,Gysbers:2019uyb}.

This paper is organized as follows. In Sec.~\ref{sec:vacuumNJL}, we briefly describe the NJL model in free space with the help of the Schwinger proper-time regularization scheme to simulate QCD confinement.
In Sec.~\ref{sec:kaonpionNJL}, we present the in-medium modifications of the pion and kaon properties that are computed within the NJL model. Section~\ref{sec:NMNJL} presents the properties of symmetric nuclear matter in the NJL model. In Sec.~\ref{sec:FFM}, we present the in-medium modifications of the kaon and pion form factors that are calculated in the NJL model. In Sec.~\ref{sec:MR}, we present the numerical results for the kaon and pion form factors in nuclear medium and free space. Summary and conclusion are devoted in Sec.~\ref{sec:summary}.

%======================================================
\section{Free space meson properties in the NJL model} \label{sec:vacuumNJL}
%======================================================
In this section, the generic expression of the three flavors NJL model Lagrangian in free space is presented, in addition to the free space quark and meson properties. All these quantities are needed to compute before computing the medium modifications of meson form factors.

%======================================================
\subsection{NJL Lagrangian}
%======================================================
The three-flavor NJL model Lagrangian in terms of the four-fermion interactions, where the gluon interaction was integrated out, is given by~\cite{Hutauruk:2016sug,Hutauruk:2018zfk,Ninomiya:2014kja,Vogl:1991qt,Buballa:2003qv}
%%%%%%%%
\begin{eqnarray}
    \label{eq:vacNJL1}
    \mathcal{L}_{\mathrm{NJL}} &=& \bar{\psi}_q \left( i \partial \!\!\!/ - \hat{m} \right) \psi_q \nonumber \\
    &+& G_\pi \Big[ \left( \bar{\psi}_q \mathbf{\lambda}_a \psi_q \right)^2 - \left( \bar{\psi}_q \mathbf{\lambda}_a \gamma_5 \psi_q \right)^2 \Big] \nonumber \\
    &-& G_V \Big[ \left( \bar{\psi}_q \mathbf{\lambda}_a \gamma^\mu \psi \right)^2 + \left( \bar{\psi}_q \mathbf{\lambda}_a 
 \gamma^\mu \gamma_5 \psi_q \right)^2 \Big],
\end{eqnarray}
where $\psi_q = (\psi_u,\psi_d, \psi_s)^T$ represents the quark field with flavor $q = (u,d,s)$ and $\hat{m} =\mathrm{diag} (m_u, m_d, m_s)$ is the current quark mass matrix. $\mathrm{\lambda}_1, \cdot \cdot \cdot, \mathrm{\lambda}_8$ are the Gell-Mann matrices in flavor space with $\lambda_0 \equiv \sqrt{\frac{1}{3}} \mathbf{1}$. It is worth noting that the four-fermion interaction term is proportional to the coupling constant $G_\pi$, which shows the direct terms of the antiquark-quark interaction in the scalar and pseudoscalar channels and has responsibility for the dynamical chiral symmetry breaking (dressed quark mass generation). Furthermore, $G_V$ is the vector coupling constant with a repulsive interaction contribution. It is also worth mentioning that the 't Hooft six-fermion (determinant) interaction term is usually considered in Lagrangian of Eq.~(\ref{eq:vacNJL1}). This term will break the global U(1)$_A$ symmetry, giving the mass splitting of $\eta$ and $\eta'$ mesons. However, in this work, we do not include such a term because it does not significantly affect the meson properties as explained in Ref.~\cite{Ninomiya:2014kja}. 

Using the mean-field approach, the dressed quark masses are determined through the quark self-energy interaction. The gap equation can be defined in the Schwinger proper-time regularization scheme and gives \cite{Hutauruk:2016sug,Ninomiya:2014kja},
%%%%%%%%%
\begin{eqnarray}
    \label{eq:vacuumNJL2}
    M_q &=& m_q + 4G_\pi i\mathrm{Tr}[S_q (p)], \nonumber \\
    &=& m_q - 4 G_\pi \langle \bar{\psi}_q \psi_q \rangle, \nonumber \\
    &=& m_q + \frac{3G_\pi M_q}{\pi^2} \int_{\tau_{\mathrm{UV}}}^{\tau_{\mathrm{IR}}} \frac{d\tau}{\tau^2} e^{-\tau M_q^2},
\end{eqnarray}
where $S_q (p) = (p\!\!\!/ + M)/(p^2-M_q^2 + i\epsilon)$ and $\langle \bar{\psi}_q \psi_q \rangle$ are, respectively, the quark propagator for $q=(u,d,s)$ and the chiral quark condensate, which is the order parameter of chiral symmetry breaking. $\tau_{\mathrm{IR}} = 1/(\Lambda_{\mathrm{IR}})^2$ and $\tau_{\mathrm{UV}} = 1/(\Lambda_{\mathrm{UV}})^2$ stand for the upper and lower limits of integration, respectively, where $\Lambda_{\rm{IR}}$ and $\Lambda_{\rm{UV}}$ are the infrared and ultraviolet cutoffs. The $\Lambda_{\rm{IR}}$ in the proper-time regularization is employed to simulate quark confinement. The $\Lambda_{\rm{UV}}$ removes the unphysical quark thresholds for the hadron decays to quarks to render the theory finite~\cite{Hellstern:1997nv}.

%======================================================
\subsection{Kaon and pion properties}
%======================================================
The kaon and pion $t$-matrices can be described by the dressed quark and dressed antiquark interaction in the kaon and pion channels within the random phase approximation (RPA), equivalently to the ladder approximation.
Summation of the bubble diagram from infinite interaction can be written by
%%%%%%%%%%
\begin{eqnarray}
    \label{eq:vacuumNJL3}
    \tau_{\pi} &=& \gamma_5 \lambda_a \Bigg( \frac{-2iG_\pi}{1 + 2 G_\pi \Pi_\pi (p^2) }\Bigg) \gamma_5 \lambda_a, \\
     \label{eq:vacuumNJL3a}
     \tau_{ K} &=& \gamma_5 \lambda_a \Bigg( \frac{-2iG_\pi}{1 + 2 G_\pi \Pi_K (p^2) }\Bigg) \gamma_5 \lambda_a,
\end{eqnarray}
where, for the pion, the sum over $a = 1,2,3$ and for the kaon $a =4,5,6,7$. The polarization insertion propagators for the pion and kaon are respectively given by 
%%%%%%%%%%
\begin{eqnarray}
    \label{eq:vacuumNJL4}
    \Pi_\pi (p^2) \delta_{ab} &=& i \int \frac{d^4k}{(2\pi)^4} \mathrm{Tr} \Big[ \gamma_5 \lambda_a S_l (p+k) \gamma_5 \lambda_b S_l (k) \Big], \nonumber \\
      \Pi_K (p^2) \delta_{ab} &=& i \int \frac{d^4k}{(2\pi)^4} \mathrm{Tr} \Big[ \gamma_5 \lambda_a S_l (p+k) \gamma_5 \lambda_b S_s (k) \Big], \nonumber \\
\end{eqnarray}
where the trace is taken for the Dirac, flavor, and color space. $S_l (k)$ and $S_s (k)$ are the light and strange quark propagators, respectively.

%======================================================
\subsection{Kaon and pion masses}
%======================================================
The kaon and pion masses can be straightforwardly determined by the pole position of the corresponding $T$ matrix in Eqs.~(\ref{eq:vacuumNJL3}) and~(\ref{eq:vacuumNJL3a}) and it gives
%%%%%%%%%%%%%
\begin{eqnarray}
    \label{eq:vacuumNJL5}
    1 + 2 G_\pi \Pi_\pi (p^2 = m_\pi^2) &=& 0, \\
    1 + 2 G_\pi \Pi_K (p^2 = m_K^2) &=& 0,
\end{eqnarray}
where near the bound state pole position of the $T$ matrix, it is given by
%%%%%%%%%
\begin{eqnarray}
    \label{eq:vacuumNJL6}
    T_{\pi,K} \simeq \gamma_5 \lambda_a \Bigg( \frac{ig_{mq}^2}{p^2 -m_k^2 + i\epsilon} \Bigg) \gamma_5 \lambda_a.
\end{eqnarray}
With $g_{mq}$ represents the meson-quark coupling constant. To determine the meson-quark coupling constant, we expand the $t-$matrix in Eq.~(\ref{eq:vacuumNJL3})at around the pole $p^2 = m_{\pi, K}^2$ and it gives
%%%%%%%%%%%
\begin{eqnarray}
    \label{eq:vacuumNJL7}
    \Pi_{\pi,K} (p^2) &=& \Pi_{\pi K} (m_{\pi K}^2) + \frac{\partial \Pi_{\pi K} (p^2)}{\partial p^2 } \Bigg|_{p^2 = m_{\pi K}^2} \nonumber \\
    &\times& (p^2 -m_{\pi K}^2) + \cdot \cdot \cdot,
\end{eqnarray}
we then can define the meson-quark coupling constant by
%%%%%%%%%
\begin{eqnarray}
    \label{eq:vacuumNJL8}
  Z_K^{-1} = \Big[ g_{\pi q q}\Big]^{-2} &=& - \frac{\partial \Pi_{\pi} (p^2)}{\partial p^2} \Bigg|_{p^2 = m_{\pi}^2}, \\
  Z_\pi^{-1} = \Big[ g_{K q q}\Big]^{-2} &=& - \frac{\partial \Pi_{K} (p^2)}{\partial p^2} \Bigg|_{p^2 = m_{K}^2}.
\end{eqnarray}

%======================================================
\subsection{Meson weak-decay constants}
%======================================================
The kaon and pion weak-decay constants are defined by the matrix element of the meson to hadron in a vacuum, which is given by $\langle 0| j_a^\mu (0) |\pi (K)_b (p) \rangle$ with $j_a^\mu$ stands for the axial-vector current operator for flavor $a$. Thus, the kaon and pion weak-decay constants are respectively given by
%%%%%%%%%%%%
\begin{eqnarray}
    \label{eq:vacuumNJL9}
    \langle 0|j_a^\mu (0) |\pi_b (p) \rangle = ip^\mu f_\pi \delta_{ab},\\
     \langle 0|j_a^\mu (0) |K_b (p) \rangle = ip^\mu f_K \delta_{ab},
\end{eqnarray}
where in the NJL model, it can be defined by
%%%%%%%%%%%%
\begin{eqnarray}
    \label{eq:vacuumNJL10}
    ip^\mu f_\pi \delta_{ab} &=& - \frac{g_{\pi q q}}{2} \int \frac{d^4k}{(2\pi)^4} \nonumber \\
    &\times& \mathrm{Tr} \Big[ \gamma^\mu \gamma_5 \lambda_a S_l(k+p) \gamma_5 \lambda_b S_l(k)\Big], \\
     ip^\mu f_K \delta_{ab} &=& - \frac{g_{K q q}}{2} \int \frac{d^4k}{(2\pi)^4} \nonumber \\
    &\times& \mathrm{Tr} \Big[ \gamma^\mu \gamma_5 \lambda_a S_l(k+p) \gamma_5 \lambda_b S_s(k)\Big]. 
\end{eqnarray}
We then apply the Feynman parameterization, and the Wick rotation and introduce the Schwinger proper-time regularization scheme. The final expression of the kaon and pion decay constants are respectively given by
%%%%%%%%%%%%
\begin{eqnarray}
    \label{eq:vaccumNJL11}
    f_K &=& \frac{3g_{Kqq}}{4\pi^2} \int_0^1 dx \int_{\tau_{\mathrm{UV}}}^{\tau_{\mathrm{IR}}} \frac{d\tau}{\tau} \Big[ M_s + x(M_l-M_s)\Big]\nonumber \\
    &\times& e^{-\tau \Big( M_s^2-x(M_s^2 -M_l^2) - x(1-x)m_K^2 \Big)},\\
       f_\pi &=& \frac{3g_{\pi qq}M_l}{4\pi^2} \int_0^1 dx \int_{\tau_{\mathrm{UV}}}^{\tau_{\mathrm{IR}}} \frac{d\tau}{\tau} e^{-\tau \Big( M_l^2 - x(1-x)m_\pi^2 \Big)}.\nonumber \\
\end{eqnarray}
Next, we will extend these free space NJL expressions into the nuclear medium in Sec.~\ref{sec:kaonpionNJL}.

%======================================================
\section{Kaon and pion properties in nuclear medium} 
\label{sec:kaonpionNJL}
%======================================================
Here, we describe the in-medium properties of the quark and kaon, which is an extension of the kaon and pion properties in Sec.~\ref{sec:vacuumNJL}. The NJL dynamical light quark masses in the nuclear medium in the Schwinger proper-time regularization scheme are given by~\cite{Whittenbury:2015ziz}
%%%%%%%%%
\begin{eqnarray}
    \label{eq:NJL1}
    M_l^* &=& m_l + 4iG_\pi \mathrm{Tr} [S^{*}_l (p^{*})], \nonumber \\
    &=& m_l + \frac{3G_\pi M_l^*}{\pi^2} \int_{\tau_{\rm{UV}}}^\infty d\tau \frac{e^{-\tau M_l^{*2}}}{\tau^2} \nonumber \\
    &-& \frac{6G_\pi M^*_l}{\pi^2} \Bigg[ \mu_q \sqrt{\mu_l^2 -M_l^{*2}} \nonumber \\
    &-& M_l^{*2} \log \left( \frac{\mu_l +\sqrt{\mu_l^2 -M^{*2}_l}}{M_l^{*}} \right) \Bigg]
\end{eqnarray}
where $M_l^{*}$ and $\mu_l$ represent the effective constituent light quark masses and the light quark chemical potential. The coupling constant of $G_\pi$ is the four-fermion coupling constant, where the values are the same as in the free space. Note that the effective mass for the strange quark is similar to that for the free space (See Eq.~(\ref{eq:vacuumNJL2})). It is worth noting that the $\tau_{\rm{IR}}$ is taken to infinity in the nuclear medium with $\Lambda_{\rm{IR}} \simeq 0$ in the deconfined phase, meaning the possibility of the nucleon decay into quarks in the unphysical threshold can occur~\cite{Bentz:2002um}. The in-medium dressed light and strange quark propagators are respectively given by~\cite{Klevansky:1992qe}
%%%%%%%%%
\begin{eqnarray}
\label{Eq:NJL2}
S_l^{*} (k^{*}) &=& \frac{k\!\!\!/^{*} + M^{*}_l}{k^{*2} -M_l^{*2}
+ i \epsilon} \nonumber \\
&+& \frac{i\pi}{E_{k,l}} \left( k\!\!\!/ + M_l^* \right) \Theta \left(\mu_l - E_{k,l} \right) \delta (k_0 -E_{k,l}), \nonumber \\
S_s^* (k^*) &=& S_s (k) = \frac{k\!\!\!/ + M_s}{k^2 - M_s^2 + i\epsilon},
\end{eqnarray}  
where $E_{k,l} = \sqrt{k^2 + M_l^{* 2}}$ with the subscript $l$ stands for the up and down quarks. The asterisk symbol denotes the nuclear medium effect, which enters the light quark momentum $k^\mu$ by $k^{*\mu} = k^\mu + V^\mu$. This is due to the vector meson mean field $V^\mu =\left( V^0,\mathbf{0} \right)$. Note that the medium modifications of the space component of the light quark momentum of $k^{*\mu}$ can be ignored since the size contribution is insignificant~\cite{Krein:1998vc}. Substituting the in-medium modifications of the quark propagator into the constituent quark mass formula in Eq.~(\ref{eq:NJL1}), we then apply the shift variable of the integral to eliminate the vector potential that enters the light quark momentum. For the strange quark, it is widely known that the strange quark is weakly coupled to the scalar and vector mean fields in the NJL nuclear matter, so it is reasonably assumed that the in-medium modifications of the strange quark propagator are similar to those in the free space, implying that the in-medium constituent quark mass is the same as that in free space ($M_s^* = M_s$).

Now we turn to present the description of the bound state of the kaon and pion, which is obtained by elucidating the Bethe-Salpeter equation (BSE) in random phase approximation (RPA). The solution of the BSE in the pion and kaon channels is determined by the two-body scattering $t$-matrix. The reduced $t$-matrices in the pion and kaon mesons and vector mesons are respectively obtained by
%%%%%%%%
\begin{eqnarray}
  \label{eq:NJL3}
    t_K^{*} (p^*) &=& \frac{-2i G_\pi}{1+ 2 G_\pi \Pi_K^{*} (p^{*2})}, \\
     t_\pi^{*} (p^*) &=& \frac{-2i G_\pi}{1+ 2 G_\pi \Pi_\pi^{*} (p^{*2})},\\
     t_V^{*\mu \nu} (p^{*}) &=& \frac{-2iG_\rho}{1 + 2 G_\rho \Pi_V^{*} (p^{*2}} \nonumber \\
     &\times& \Bigg[ g^{\mu \nu} + 2 G_\rho \Pi^{*}_V (p^{*2}) \frac{p^{*\mu} p^{*\nu}}{p^2}\Bigg],
\end{eqnarray}  
where $G_\rho$ is the four-fermion coupling constant for the vector meson channels. It is worth noting that for simplicity, in the present work, we choose the vector coupling $G_\rho = G_V$~\cite{Hell:2014xva,Klimt:1990ws}. The polarization insertions or the bubble diagrams in the nuclear medium can be introduced by
%%%%%%%%%
\begin{eqnarray} 
    \label{eq:NJL4} 
    \Pi^{*}_K (p^{*2}) &=& 6i \int \frac{d^4k}{(2\pi)^4} \mathrm{Tr} \Big[ \gamma_5 S_l^* (k^*) \gamma_5 S_s^{*} (k^{*}+p^{*} \Big],\nonumber \\
    \Pi^{*}_\pi (p^{*2}) &=& 6i \int \frac{d^4k}{(2\pi)^4} \mathrm{Tr} \Big[ \gamma_5 S_l^* (k^*) \gamma_5 S_l^{*} (k^{*}+p^{*} \Big],\nonumber \\
    \Pi_V^{ll} (p^{*2}) P_T^{\mu \nu} &=& 6i \int \frac{d^4k}{(2\pi)^4} \mathrm{Tr} \Big[ \gamma^\mu S_l^* (k^*) \gamma^\nu S_l^* (k^* + p^*)\Big], \nonumber \\
\end{eqnarray}    
where $P_T^{\mu \nu} = g^{\mu \nu} - q^\mu q^\nu/q^2$, $\mathrm{Tr}$ are only included for the Dirac indices. It is worth noting that the polarization insertion for the light quarks, $\Pi_\rho^* = \Pi_\omega^* = \Pi_V^{*ll}$, while for the strange quarks, $\Pi^{*}_\phi = \Pi_V^{*ss}$.

After extracting the polarization insertion for the pion and kaon, we can also straightforwardly determine the in-medium meson masses for the kaon and pion from the pole position of the corresponding $t$-matrix, which are given by
%%%%%%%%%%
\begin{eqnarray}
    \label{eq:NJL5}
    1 + 2 G_\pi \Pi^{*}_K (p^{*2} = m_K^{*2}) &=& 0,\\
    1 + 2 G_\pi \Pi^{*}_\pi (p^{*2} = m_K^{*2}) &=& 0.
\end{eqnarray}

Next, using the polarization insertions of the kaon and pion, other properties, such as the kaon-quark coupling constant in the medium and pion-quark coupling constant in the nuclear medium, can be easily computed by taking the first derivative of the polarization insertion of the kaon and pion, respectively, with $p^{*2} = m_K^{*2}$. Mathematically, it can be formulated by
%%%%%%%%%%
\begin{eqnarray}
    \label{eq:NJL6} 
\left[Z_K^{*}\right]^{-1} =\Big[ g_{K q q}^{*} \Big]^{-2} &=& - \frac{\partial \Pi_K^{*} (p^{*2})}{\partial p^{*2}} \Bigg|_{p^{*2} = m_K^{*2}},\\
\left[Z_\pi^{*}\right]^{-1} =\Big[ g_{\pi q q}^{*} \Big]^{-2} &=& - \frac{\partial \Pi_\pi^{*} (p^{*2})}{\partial p^{*2}} \Bigg|_{p^{*2} = m_\pi^{*2}}. 
\end{eqnarray}

%======================================================
\section{Nuclear matter in the NJL model} \label{sec:NMNJL}
%======================================================
In this section, we present the symmetric nuclear matter NJL model built at the quark level, which is adapted from Ref.~\cite{Bentz:2001vc}. After applying the quark bilinear into NJL Lagrangian in Eq.~(\ref{eq:vacNJL1}), the NJL effective Lagrangian in the symmetric nuclear matter (NM) is given by~\cite{Bentz:2001vc}
%%%%%%%%%%
\begin{eqnarray} 
\label{eq:NJL7}
\mathcal{L}_{\mathrm{NM-NJL}} &=& \bar{\psi} \left( i \partial\!\!\!/ - M - V\!\!\!\!/ \right)\psi - \frac{\left( M-m\right)^2}{4G_\pi}\nonumber \\
&+& \frac{V_\mu V^\mu}{2G_\omega} + \mathcal{L}_{I},
\end{eqnarray}  
where $M = M_l$, which is the light constituent quark mass. In this work, we consider SU(2) isospin symmetry that gives $ M_u = M_d = M$. $\mathcal{L}_{I}$ represents the interaction Lagrangian. Using Fierz transformation, the $\mathcal{L}_{I}$ in Eq.~(\ref{eq:NJL7}) can be rewritten as a sum of $qq$ interactions of the isoscalar-scalar ($0^+, T=0$) and isovector-axial vector ($1^+, T=1$) and the NJL interaction Lagrangian takes the form~\cite{Bentz:2001vc,Ishii:1995bu}
%%%%%%%%%%%%%
\begin{eqnarray}
    \label{eq:NJL7a}
    \mathcal{L}_{I,qq} &=& G_s \left[ \bar\psi_q \gamma_5 C \tau_2 \beta_A \bar\psi_q^T \right] \left[ \psi^T C^{-1} \gamma_5 \tau_2 \beta_A \psi\right] \nonumber \\
    &+& G_a \left[ \bar\psi_q \gamma_\mu C \tau_i \tau_2 \beta_A \bar\psi_q^T\right] \nonumber \\
    &\times& \left[ \psi_q^T C^{-1} \gamma^\mu \tau_2 \tau_i \beta_A \psi_q \right],
\end{eqnarray}
%%%%%%%%%%%%%
where $C=i \gamma_2 \gamma_0$ is the charge conjugation matrix, $\beta_A = \sqrt{\frac{3}{2}} \lambda_A$ with $A = 2,5,7$ and the couplings $G_s$ and $G_a$ are respectively the strength of the scalar and axial vector $qq$ interactions. Note that the coupling $G_s$ is determined to reproduce the free nucleon mass. In nuclear matter, the repulsive $G_V$ coupling constant in the original NJL Lagrangian is replaced by the repulsive $G_\omega$ coupling constant, representing the nucleon-nucleon interaction by exchanging $\omega$ vector-meson mean field which couples to quarks.

Using the standard hadronization technique, the expression of the effective potential for symmetric nuclear matter can be determined from the NJL Lagrangian. In the mean-field approximation, the energy density for NM is given by~\cite{Bentz:2001vc}
%%%%%%%%%
\begin{eqnarray}
\label{eq:NJL9} 
\mathcal{E} = \mathcal{E}_V - \frac{V_0^2}{4G_\omega} + 4 \int \frac{d^3p}{(2\pi)^3} \Theta (p_F - |\mathbf{p}|)\epsilon_N,
\end{eqnarray}  
where $\epsilon_N = \sqrt{M_N^{2} + \mathbf{p}^2} + 3V_0 \equiv E_N + 3V_0$, where $M_N(M)$ is the nucleon mass in medium obtained from the pole of the quark-diquark $T$-matrix and the nucleon Fermi momentum can be defined by $p_F = \sqrt{(\mu_N-3V_0)^2 - M_N^2}$, where $\mu_N$ is nucleon chemical potential. The last term is the Fermi motion of nucleons in the scalar and vector mean fields that couple to the quarks. For further details of the derivations and explanations, we refer to Ref.~\cite{Bentz:2001vc}. The vacuum contribution of the quark loop can be defined by
%%%%%%%
\begin{eqnarray}
\label{eq:NJL10}    
\mathcal{E}_V &=& 12i \int \frac{d^4k}{(2\pi)^4} \ln \Bigg( \frac{k^2 -M^2 + i\epsilon}{k^2 - M_0^2 + i \epsilon}\Bigg) + \frac{\left( M-m\right)^2}{4G_\pi} \nonumber \\
&-& \frac{\left( M_0-m\right)^2}{4G_\pi},
\end{eqnarray}  
where $M_0 $ is the vacuum value of the constituent quark mass at zero baryon density. 
The isoscalar-vector mean field and the constituent quark mass can be respectively defined by
%%%%%%%%%
\begin{eqnarray}
\label{eq:NJL8}
V^\mu &=& 2 G_\omega \langle \rho|\bar{\psi}_q \gamma^\mu \psi_q |\rho \rangle = 2 \delta^{0\mu} G_\omega \langle \psi_q^\dagger \psi_q \rangle, \\
\label{eq:NJL8a}
M &=& m - 2 G_\pi \langle \rho| \bar{\psi}_q \psi_q |\rho\rangle,
\end{eqnarray}  
where the vector potential is defined by $V=(V_0,\mathbf{0})$. Next, using the minimum condition $\partial \mathcal{E} / \partial V_0 =$ 0, the value of $V_0$ is obtained as
%%%%%%%%%%
\begin{eqnarray}
\label{eq:NJL11}    
    V_0 &=& 6 G_\omega \rho_B,
\end{eqnarray}  
where $\rho_B = 2 p_F^3 /3 \pi^2$ represents the baryon density with $p_F$ is the Fermi momentum of the baryon. It is worth noting that the constituent quark mass for fixed baryon density must satisfy the minimum condition $\partial \mathcal{E} /\partial M =$ 0 to give a similar expression of the in-medium constituent quark mass in Eq.~(\ref{eq:NJL8a}).

Using the expression of the energy density in Eq.~(\ref{eq:NJL9}), we determine the binding energy per nucleon for symmetric nuclear matter, which is given by
%%%%%%%%
\begin{eqnarray}
\frac{E_B}{A} &=& \frac{\mathcal{E}}{\rho_B} - M_{\mathrm{N0}},
\end{eqnarray}  
where $M_{\mathrm{N0}}$ stands for the free space nucleon mass. Our symmetry nuclear matter NJL model reproduces well the binding energy per nucleon $E_B/A = -$15.7 MeV at saturation density $\rho_0 =$ 0.16 fm$^{-3}$. 

Using the quark properties calculated in the nuclear matter NJL model, we obtain the quantities such as dressed quark masses, kaon and pion masses, pion and kaon quark coupling constants, and the quark condensates in the nuclear medium. A summary of these in-medium modification quantities is tabulated in Table.~\ref{tab:NJL2}. Next, these quantities are needed to compute the kaon and pion electromagnetic form factors in a nuclear medium. 
%%%
\begin{table*}[t]
	\begin{ruledtabular}
		\renewcommand{\arraystretch}{1.1}
		\caption{The in-medium modifications of the constituent quark mass, kaon mass, kaon decay constant, the constituent quark mass, pion mass, pion decay constant, quark condensate, and effective nucleon mass calculated in the NJL model. The units are in MeV. Note that $-\langle \bar{s} s \rangle^{*1/3} =$ 0.150 GeV and unit of $\langle \bar{u} u \rangle^{*1/3}$ is in GeV.}
		\label{tab:NJL2}
		\begin{tabular}{cccccccccccc}
		  $\rho_B/\rho_0$  & $M_q^*$ & $M_s^*$ & $m_{\rm K}^*$ & $f_K^*$ &$g_{\rm K qq}^{*}$ & $m_{\rm \pi}^*$ & $f_\pi^*$ &$g_{\rm \pi qq}^*$ &  $-\langle \bar{u}u \rangle^{*1/3}$ & $M_N^*$ &$g_A^{*}/g_A $ \rule[-2ex]{0pt}{4ex} \\ 
\hline  
		 $0.00$ &  400 & 611 & 495  & 97 & 4.572 & 140  & 93 & 4.225 & 0.171 & 934.36 &1.00\\
	$0.25$ &  379 & 611 & 495  & 99 & 4.467 & 137  &92 & 4.040 & 0.168 & 887.13 & 0.991 \\
		 $0.50$ &  360 & 611 & 496  & 100 & 4.375 & 136  & 91 & 3.882 & 0.165 & 842.54 &0.992 \\
		 $0.75$ &  343 & 611 & 499  & 101 & 4.296 & 134  & 90 & 3.750 & 0.162 & 806.43 & 0.995 \\
		 $1.00$ &  328 & 611 & 501  & 102 & 4.224 & 133  & 89 & 3.639 & 0.160 & 778.03 & 0.998\\
		 $1.25$ &  316 & 611 & 504  & 103 & 4.170 & 132  & 87 & 3.555 & 0.158 & 756.01 & 0.989\\
		 $1.50$ &  306 & 611 & 506  & 104 & 4.123 & 132  & 86 & 3.488 & 0.156 & 738.98 & 0.991\\
		 $1.75$ &  297 & 611 & 509  & 104 & 4.084 & 132  & 85 & 3.429 &0.154 & 725.73 & 0.992\\
          $2.00$ &  289 & 611 & 511  & 105 & 4.047 & 132  & 84 & 3.379 &  0.153 & 715.34 & 0.993
		\end{tabular}
		\renewcommand{\arraystretch}{1}
	\end{ruledtabular}
\end{table*}
%%%

%======================================================
\section{Nuclear medium form factors} \label{sec:FFM}
%======================================================
Here we present the electromagnetic form factor for the kaon and pion in the nuclear medium. The electromagnetic form factors can be extracted from the matrix element of the electromagnetic current, which is given by~\cite{Hutauruk:2016sug}
%%%%%%%%%%
\begin{eqnarray}
\label{eq:NJL12}    
    \mathcal{J}^\mu (p^{'*},p^{*}) &=& \left( p^{'*\mu} + p^{*\mu} \right) F_K^{*} (Q^2),
\end{eqnarray}  
where $F_K^{*} (Q^2)$ is the medium modification of the kaon form factor and it can be written in terms of the quark sector form factors,
%%%%%%%%%
\begin{eqnarray}
\label{eq:NJL13}
F_{K}^* (Q^2) &=& e_u F_K^{*u} (Q^2) + e_d F_K^{*d} (Q^2) + e_s F_K^{*s} (Q^2) + \cdot \cdot \cdot . \nonumber \\
\end{eqnarray}  
Analogously, it can be applied to the pion case by replacing the strange quark with the light quark. 

Following the free space calculation~\cite{Hutauruk:2016sug}, the in-medium modifications of the kaon and pion form factors can be computed from the sum of two Feynman diagrams, as in Ref.~\cite{Hutauruk:2016sug}. It is worth noting that we consider the in-medium dressed quark-photon vertex $\Lambda_{\gamma q}^{*\mu} = \gamma^\mu F_{1Q}^{*} (Q^2)$, rather than in medium bare quark-photon vertex $\Lambda_{\gamma q}^{*\mu}  = \hat{Q} \gamma^\mu$. This form satisfies the Ward-Takahashi identity (WTI) defined by
%%%%%
\begin{eqnarray}
\label{eq:NJL14}
    q_\mu \Lambda_{\gamma Q}^{*\mu} (p^{'*},p^*) &=& e_{Q} \Big[ S_Q^{*-1} (p^{'*}) - S_{Q}^{*-1} (p^*) \Big].
\end{eqnarray}  

The dressed quark form factors for up, down, and strange quarks in the medium are respectively defined by~\cite{Hutauruk:2016sug}
%%%%%%
\begin{eqnarray}
    \label{eq:NJL14a}
    F_{1U}^{*} (Q^2) &=& \frac{e_u}{1 +2G_\rho \Pi_V^{*ll} (Q^2)}, \\
     F_{1D}^{*} (Q^2) &=& \frac{e_d}{1 +2G_\rho \Pi_V^{*ll} (Q^2)},\\
      F_{1S}^{*} (Q^2) &=& \frac{e_s}{1 +2G_\rho \Pi_V^{*ss} (Q^2)},
\end{eqnarray}
where the in-medium bubble diagrams are the same as defined in Eq.~(\ref{eq:NJL4}). 

After performing the calculation such as Feynman parameterization and applying the Schwinger proper-time regularization scheme, the final expression for the in-medium kaon and pion form factors are respectively given by~\cite{Hutauruk:2016sug}
%%%%%%%%%%
\begin{eqnarray}
    \label{eq:NJL15}
    F_K^* (Q^2) &=& F_{1U}^* (Q^2) f_K^{*ls} (Q^2) - F_{1S}^{*} (Q^2) f_K^{*sl} (Q^2), \nonumber \\
     F_\pi^* (Q^2) &=& F_{1U}^* (Q^2) f_\pi^{*ll} (Q^2) -F_{1D}^{*} (Q^2) f_\pi^{*ll} (Q^2), \nonumber \\
\end{eqnarray}
where the $f_K^{*ls} (Q^2)$ and $f_\pi^{*ll} (Q^2)$ are respectively the bare kaon and pion form factors in the nuclear medium.

%======================================================
\section{Numerical Result} \label{sec:MR}
%======================================================
In this section, we present the numerical results for the kaon and pion properties and their elastic and quark sector form factors in the nuclear medium. The NJL parameter used in this present work is $G_\pi$, $G_\omega$, $G_\rho$, $G_a$, and $G_s$, as used in Refs.~\cite{Hutauruk:2021kej,Hutauruk:2016sug,Tanimoto:2019tsl,Bentz:2001vc}. In this work, the regularization parameters of the $\Lambda_{\mathrm{IR}}$ in the Schwinger proper-time regularization scheme are fixed to $\Lambda_{\mathrm{\rm{IR}}} =$ 240 MeV, which is set the same as the order of $\Lambda_{\mathrm{QCD}}$ and the constituent quark mass at zero baryon density is $M_0 =$ 400 MeV. The $G_\pi$ and $\Lambda_{\rm{UV}}$ parameters are determined to fit the pion physical mass $m_\pi =$ 140 MeV, and pion decay constant $f_\pi =$ 93 MeV. Kaon physical mass $m_K =$ 495 MeV and kaon decay constant are used to determine the strange quark masses and kaon-quark coupling. The coupling $G_s$ is determined to reproduce the nucleon mass $M_N = M_{N0} =$ 940 MeV, and we use $\rho$-meson mass $m_\rho =$ 770 MeV. 

The fitted results give ultraviolet cutoff $\Lambda_{\mathrm{UV}} =$ 645 MeV, pion coupling constant $G_\pi =$ 19.04 GeV$^{-2}$, axial-vector diquark coupling constant $G_a =$ 2.8 GeV$^{-2}$, scalar diquark coupling constant $G_s =$ 7.49 GeV$^{-2}$, and strange constituent quark mass $M_s =$ 611 MeV. The $\phi$ meson mass is $m_\phi =$ 1001 MeV, scalar diquark mass $M_{sa} =$ 687 MeV, and axial diquark mass $M_a =$ 1027 MeV. It is worth noting that the $G_s$ and $G_a$ are determined using the Faddeev equation to produce the vacuum nucleon mass $M_N$ and axial coupling constant of nucleon $g_A =$ 1.267, as used in Ref.~\cite{Tanimoto:2019tsl,Bentz:2001vc}. The coupling constant $G_\omega$ is determined by fitting the binding energy per nucleon $E_B/A = -$15.7 MeV at saturation baryon density $\rho_0 =$ 0.16 fm$^{-3}$, it gives $G_\omega =$ 6.03 GeV$^{-2}$. The current quark mass for the up quark is given by $m_u =$ 16 MeV and the current quark mass for the strange quark is obtained  $m_s =$ 356 MeV, as used in Ref.~\cite{Hutauruk:2021kej,Hutauruk:2016sug,Tanimoto:2019tsl,Bentz:2001vc}.

%======================================================
\subsection{Meson properties in the nuclear medium}
%======================================================
%%%% FIG 1%%%%%%%%%%%%%%%%%%%%%%%%%%
\begin{figure}[t]
    \centering
    \includegraphics[width=1.01\columnwidth]{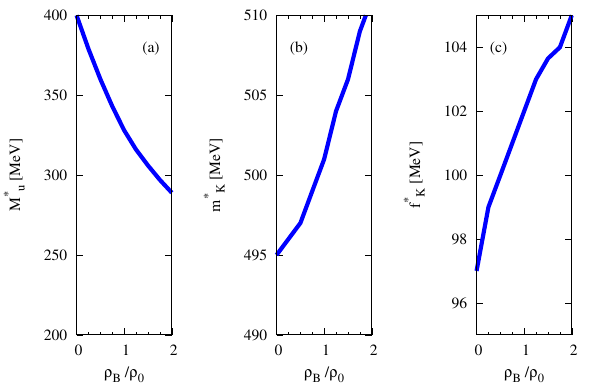}
    \caption{Quark and kaon properties as a function of $\rho_B/\rho_0$, (a) effective constituent quark mass of the up quark, (b) effective kaon mass, and (c) medium kaon decay constant.}
    \label{fig1}
\end{figure}
\begin{figure}[t]
    \centering
    \includegraphics[width=1.01\columnwidth]{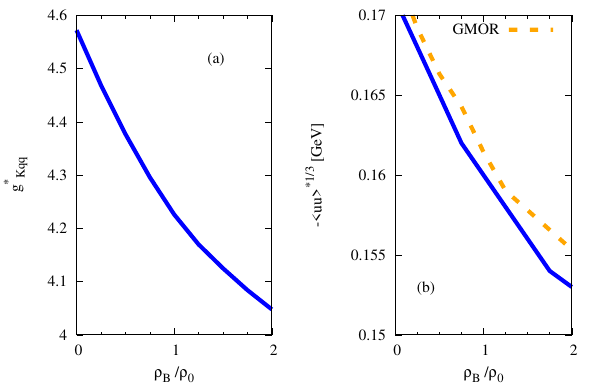}
    \caption{(a) Kaon-quark coupling constant as a function of $\rho_B/\rho_0$ and (b) quark condensate in the nuclear medium as a function of $\rho_B/\rho_0$, in comparison with that calculated via GMOR (dashed line).}
    \label{fig1b}
\end{figure}

%%%%%%%%%%%%%%%%%%%%%%%%%%%%%%%%%%%%%%
Figure~\ref{fig1}(a)-(c) shows the results for the in-medium modifications of the constituent mass of the up-quark, kaon mass, and kaon weak-decay constant, respectively. Figure~\ref{fig1}(a) indicates that the in-medium modifications of the constituent quark mass decrease as the density increases. This result is consistent with the result obtained in Ref.~\cite{Hutauruk:2018qku}. Relative to the vacuum values, the reduction rate of $M_u^*$ is approximately 18\% at $\rho_B = \rho_0$. This indicates that the chiral symmetry breaking is restored at higher baryon density approaching the current quark mass value. As a consequence, the dynamics of the valence up-quark inside the kaon increases because it has more space region to move. This results in an increase of the charge radius and kaon mass. In addition, a clear indication of the chiral symmetry-breaking restoration can also be seen from the result of the quark condensate in Fig.~\ref{fig1b}(b). The parameterization for the in-medium modifications of the constituent mass of the up-quark can be expressed as 
%%%%%%%%
\begin{eqnarray}
    \frac{M_u^*}{M_u} = 1 -0.22 \Bigg(\frac{\rho_B}{\rho_0} \Bigg) + 0.04\Bigg(\frac{\rho_B}{\rho_0} \Bigg)^2.
\end{eqnarray}

In Fig.~\ref{fig1}(b), we show the results for the in-medium modifications of kaon mass as a function of the baryon density. The effective kaon mass slowly increases with the density. The increase of kaon mass in the nuclear medium is relatively small, which is about 1.2\% at $\rho_B = \rho_0$ relative to that in vacuum (zero density). Our prediction for the effective positively charged kaon mass in the nuclear medium is consistent with the theoretical results obtained in Refs.~\cite{Waas:1996fy,Bernard:1987sx}, where they found that the kaon mass for positively charged increases as the density increases, while for the kaon negatively charged it decreases as the density increases. Here we also parameterize the kaon mass in the nuclear medium and it gives
%%%%%%%%%
\begin{eqnarray}
    \frac{m_K^*}{m_K} &=& 1 + 0.009 \Bigg( \frac{\rho_B}{\rho_0}\Bigg) +  0.04 \Bigg( \frac{\rho_B}{\rho_0}\Bigg)^2.
\end{eqnarray}

Results for the kaon weak-decay constant in the nuclear medium are shown in Fig.~\ref{fig1}(c). Similarly, with the in-medium modifications of the kaon mass, the in-medium modifications of the kaon weak-decay constant enhance with increasing densities. The increasing rate at $\rho_B = \rho_0$ is approximately about 5.2\% relative to that in vacuum. The parameterization for the in-medium modifications of the kaon weak-decay constant is given by
%%%%%%%%
\begin{eqnarray}
    \frac{f_K^*}{f_K} &=& 1 + 0.06 \Bigg( \frac{\rho_B}{\rho_0}\Bigg) - 0.01\Bigg( \frac{\rho_B}{\rho_0}\Bigg)^2.
\end{eqnarray}

Next, we also compute the kaon-quark coupling constant in the nuclear medium and free space (at zero density). For the in-medium modifications of the kaon-quark coupling constant, it is found that the coupling constant decreases as the density increases, as shown in Fig.~\ref{fig1b} (a). The decreasing rate of the in-medium kaon-quark coupling constant at $\rho_B = \rho_0$ is about 7.6\% relative to that in vacuum. The parameterization for the in-medium modifications of the kaon-quark coupling constant is given by
%%%%%%%%
\begin{eqnarray}
    \frac{g_{Kqq}^*}{g_{K qq}} &=& 1 - 0.09 \Bigg( \frac{\rho_B}{\rho_0}\Bigg) + 0.02\Bigg( \frac{\rho_B}{\rho_0}\Bigg)^2.
\end{eqnarray}

Results for $-\langle \bar{u}u \rangle^{1/3}$ as a function of $\rho_B/\rho_0$ are depicted in Fig.~\ref{fig1b}(b). As expected, the chiral condensate in the nuclear medium decreases as the density increases. At $\rho_B = \rho_0$ the in-medium modifications reduce the chiral condensate value by about $6.4 \%$, relative to its value at $\rho_B = 0$ (vacuum). Then, by recalling that the quark condensate represents an order parameter for the chiral transition one may conclude that the behavior observed in Fig.~\ref{fig1b}(b) is in line with the expected (partial)
restoration of chiral symmetry at high chemical potential values~\cite{Buballa:2003qv,Klevansky:1992qe,Asakawa:1989bq}. We then do a parameterization for the in-medium modifications of the quark condensate and it has a form
%%%%%%%%%%%%%%%%%%%%%%%%%
\begin{eqnarray}
    \frac{\langle \bar{u} u \rangle^*}{\langle \bar{u} u \rangle} &=& 1 - 0.08 \Bigg( \frac{\rho_B}{\rho_0}\Bigg) + 0.01\Bigg( \frac{\rho_B}{\rho_0}\Bigg)^2.
\end{eqnarray}

%%%% FIG 1%%%%%%%%%%%%%%%%%%%%%%%%%%
%%%%%%%%%%%%%%%%%%%%%%%%%%%%%%%%%%%%%%

In Fig.~\ref{fig1b}(b), we also show the results for the quark condensate (dashed line) computed through the Gell-Mann-Oakes-Renner relation (GMOR), which is expressed as 
%%%%%%%
\begin{eqnarray}
\label{eq:GMOR}     
    f_\pi^2 m_\pi^2 &=& - \frac{1}{2} (m_u + m_d) \langle \bar{u}u + \bar{d}d \rangle.
\end{eqnarray}
By inserting the values of $f_\pi$ and $m_\pi$ given in Table~\ref{tab:NJL2}, along with $m_u=m_d =$ 16.4 MeV, to Eq.~(\ref{eq:GMOR}), the quark condensate for various density can be obtained. Figure~\ref{fig1b}(b) clearly shows that the results for the quark condensate from the GMOR in Eq.~(\ref{eq:GMOR}) have a similar tendency compared to that calculated directly in the NJL model. This explicitly indicates that the NJL model with the Schwinger proper-time regularization scheme is consistent with the chiral symmetry in QCD. 
Another interesting result is at zero nuclear matter density, the NJL model satisfies the chiral limit $m_q =0$, where $m_q$ and pion mass $m_\pi$ vanish, the $M_q \neq 0$ since the chiral condensate does not vanish.

%%%% FIG 2%%%%%%%%%%%%%%%%%%%%%
\begin{figure}[t]
    \centering
    \includegraphics[width=1.0\columnwidth]{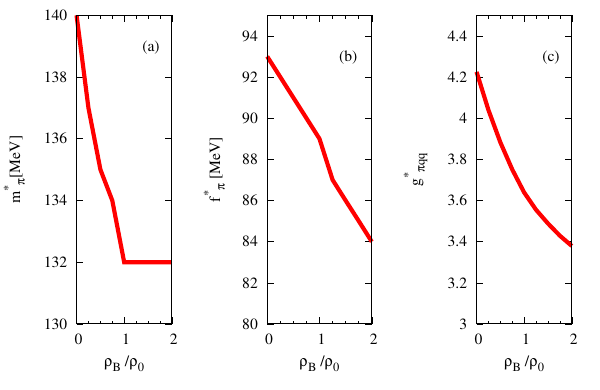}
    \caption{Pion properties as a function of $\rho_B/\rho_0$, (a) effective pion mass, (b) medium pion weak-decay constant, and (c) medium pion-quark coupling constant.}
    \label{fig2}
\end{figure}
%%%%%%%%%%%%%%%%%%%%%%%%%%%%%%%%%%%%%

Results for the effective pion mass are shown in Fig.~\ref{fig2}(a). It clearly shows that the effective pion mass decreases as density increases up to $\rho_B = 0.75 \rho_0$ and at a large density of $\rho_B > 1.00 \rho_0$, the values of the pion mass are almost unchanged, which is in a good agreement with the result given in Ref.~\cite{Bernard:1987sx}. The differences between the effective pion mass and the free pion mass are approximately within 5.7\% at $\rho_0$. The parameterization form for the in-medium modifications of the pion mass is given by
%%%%%%%%
\begin{eqnarray}
    \frac{m_\pi^*}{m_\pi} = 1 -0.08 \Bigg(\frac{\rho_B}{\rho_0} \Bigg) + 0.02\Bigg(\frac{\rho_B}{\rho_0} \Bigg)^2.
\end{eqnarray}

Results for the in-medium pion weak-decay constant are shown in Fig.~\ref{fig2}(b). The in-medium modifications of the pion weak-decay constant decrease as the baryon density increases, which is rather different from the results for the in-medium modifications of the kaon weak-decay constant, as shown in Fig.~\ref{fig1}(c). The suppression of the pion weak-decay constant at $\rho_B = \rho_0$ is approximately about 4.3\% relative to that in vacuum. The parameterization form for the in-medium modifications of the pion mass is given by
%%%%%%%%
\begin{eqnarray}
    \frac{f_\pi^*}{f_\pi} = 1 -0.05 \Bigg(\frac{\rho_B}{\rho_0} \Bigg) - 0.002\Bigg(\frac{\rho_B}{\rho_0} \Bigg)^2.
\end{eqnarray}

Moreover, we also numerically compute the pion-quark coupling constant in the nuclear medium. Our results for the in-medium modifications of the pion-quark coupling constant are shown in Fig.~\ref{fig2}(b). It clearly shows that the in-medium modifications of the pion-quark coupling constant decrease with increasing baryon densities. The decreasing rate is approximately about 13.9\% at $\rho_B = \rho_0$ relative to that in vacuum. The parameterization form for the in-medium modifications of the pion-quark coupling constant is given by
%%%%%%%%
\begin{eqnarray}
    \frac{g_{\pi q q}^*}{g_{\pi q q}} = 1 -0.17 \Bigg(\frac{\rho_B}{\rho_0} \Bigg) + 0.04\Bigg(\frac{\rho_B}{\rho_0} \Bigg)^2.
\end{eqnarray}

Using the results for the pion properties in the nuclear medium, we can straightforwardly calculate the quenching of the nuclear weak axial-vector coupling constant through the Goldberger-Treiman relation (GTR). In the nuclear medium, the GTR at the quark level can be simply written as~\cite{Ramalho:2012pu}
%%%%%%%%%%
\begin{eqnarray}
\frac{g_A^*}{g_A} &=& \Bigg( \frac{g_{\pi qq}^*}{g_{\pi qq}}\Bigg) \Bigg( \frac{f_\pi^*}{f_\pi}\Bigg) \Bigg(\frac{M_q}{M_q^*} \Bigg),
\end{eqnarray}  
where $M_N^*$ and $M_N$ are, respectively, the effective and vacuum nucleon masses. The values of $M_N^*$ calculated in the NJL model for various densities are listed in Table~\ref{tab:NJL2}. Parameterization of the in-medium modifications of the $g_A$ is given by
%%%%%%%
\begin{eqnarray}
\frac{g_A^*}{g_A} &=& 1 - 0.007 \Bigg( \frac{\rho_B} {\rho_0} \Bigg) +0.002\Bigg( \frac{\rho_B}{\rho_0}\Bigg)^2.
\end{eqnarray}

\begin{figure}[t]
    \centering
    \includegraphics[width=1\columnwidth]{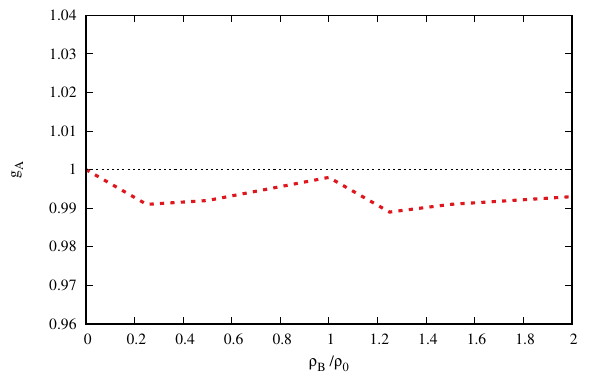}
    \caption{Quenching factor of the axial-vector coupling constant as a function of $\rho_B/\rho_0$.}
    \label{fig3}
\end{figure}

Finally, the results for the in-medium modifications of the axial-vector coupling constant for various densities are shown in Fig.~\ref{fig3}. It clearly shows that the quenching factor $q=g_A^*/g_A$  monotonically decreases as the nuclear matter density increases. By taking an experimental value of $g_A = 1.267$ in free space, at baryon density $\rho_B = \rho_0$, the reduction of $g_A$ is approximately about 0.2\% relative to that in free space, indicating very small quenching factor. This result is consistent with the results obtained in Refs.~\cite{Ma:2020tsj,Suhonen:2017krv,Dominguez:2023bjb}. The values of $g_A^*/g_A$ are also given in Table.~\ref{tab:NJL2}.

%%%% FIG 3%%%%%%%%%%%%%%%%%%%%%
%%%%%%%%%%%%%%%%%%%%%%%%%%%%%%%%%%%%%
In the next section, we will use the results of the density-dependent pion and kaon properties as input in the computation of the in-medium modifications of the pion and kaon electromagnetic form factors.

%======================================================
\subsection{Meson structure in the nuclear medium}
%======================================================
Here, numerical results for the in-medium modifications of the elastic form factors and charge radius of the pion and kaon in free space and nuclear medium calculated in the NJL model are presented.

%======================================================
\subsubsection{Elastic form factors}
\begin{figure}[t]
    \centering
    \includegraphics[width=1\linewidth]{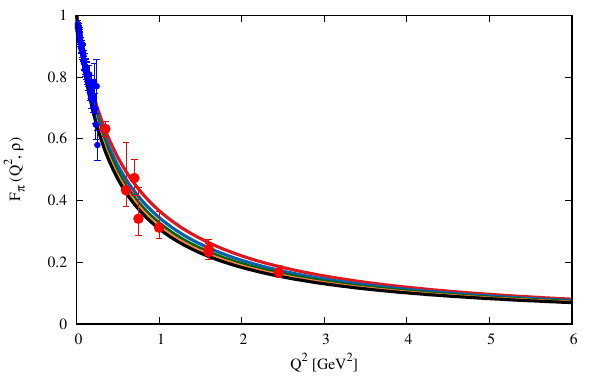}
    \caption{Pion form factor as a function of $Q^2$ for various nuclear matter densities. The red, blue, green, orange, and black lines represent $\rho_B/\rho_0 =$ 0.0 (free space), $\rho_B/\rho_0 =$ 0.50, $\rho_B/\rho_0 =$ 1.00, $\rho_B/\rho_0 =$ 1.50, and $\rho_B/\rho_0 =$ 2.00, respectively. Experimental data are taken from Refs.~\cite{Amendolia:1984nz,JeffersonLab:2008jve}.}
    \label{fig4}
\end{figure}

\begin{figure}[t]
    \centering
    \includegraphics[width=1\linewidth]{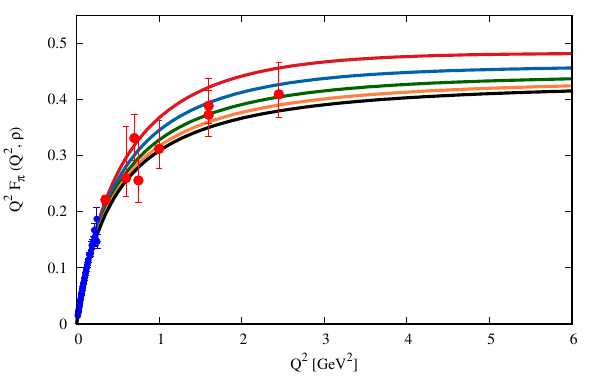}
    \caption{Same as in Fig.~\ref{fig4}, but for $F_\pi (Q^2)$ multiplied by $Q^2$.}
    \label{fig5}
\end{figure}

\begin{figure}[t]
    \centering
    \includegraphics[width=1\linewidth]{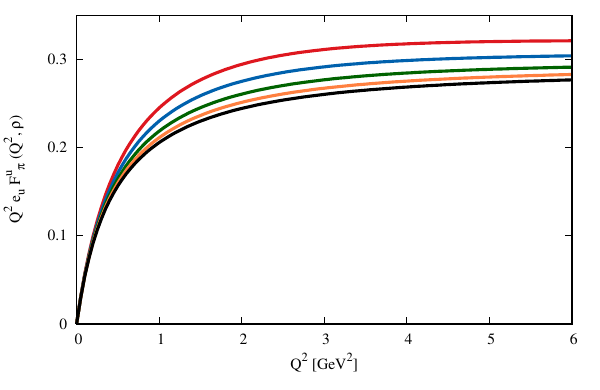}
    \caption{Same as in Fig.~\ref{fig4}, but for $e_u F^u_\pi (Q^2)$ multiplied by $Q^2$.}
    \label{fig6}
\end{figure}

\begin{figure}[t]
    \centering
    \includegraphics[width=1\linewidth]{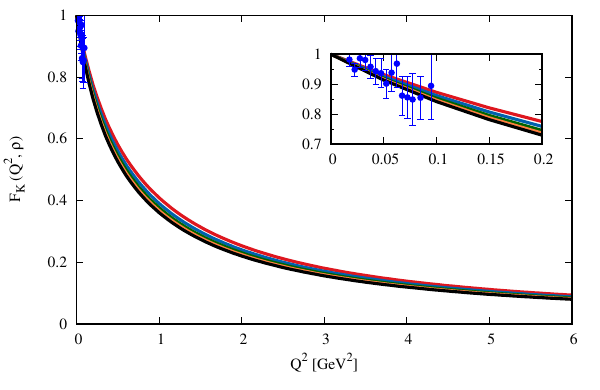}
    \caption{Same as in Fig.~\ref{fig4}, but for the kaon. Here, the experimental data are taken from Ref.~\cite{Amendolia:1986ui}.}
    \label{fig7}
\end{figure}

\begin{figure}[t]
    \centering
    \includegraphics[width=1\linewidth]{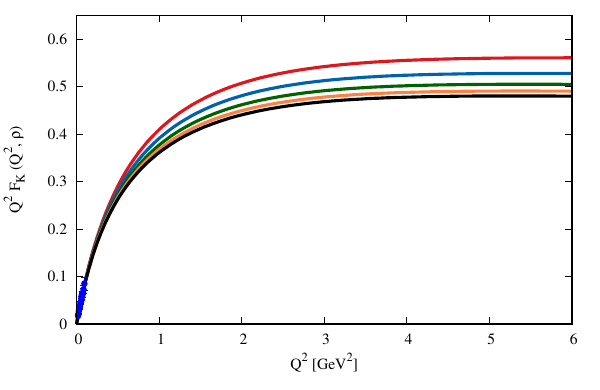}
    \caption{Same as in Fig.~\ref{fig7}, but for $F_K (Q^2,\rho_B)$ multiplied by $Q^2$.}
    \label{fig8}
\end{figure}

\begin{figure}[t]
    \centering
    \includegraphics[width=1\linewidth]{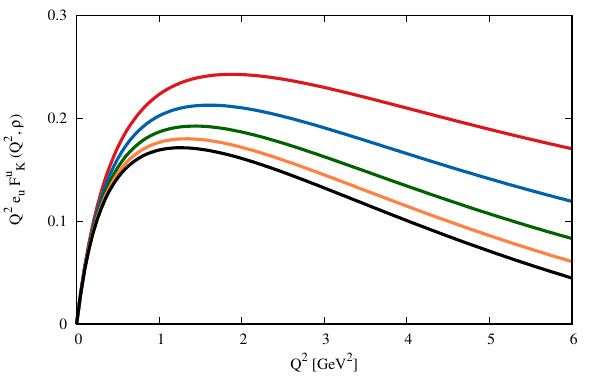}
    \caption{Same as in Fig.~\ref{fig7}, but for $e_u F^u_K (Q^2, \rho_B)$ multiplied by $Q^2$.}
    \label{fig9}
\end{figure}

\begin{figure}[t]
    \centering
    \includegraphics[width=1\linewidth]{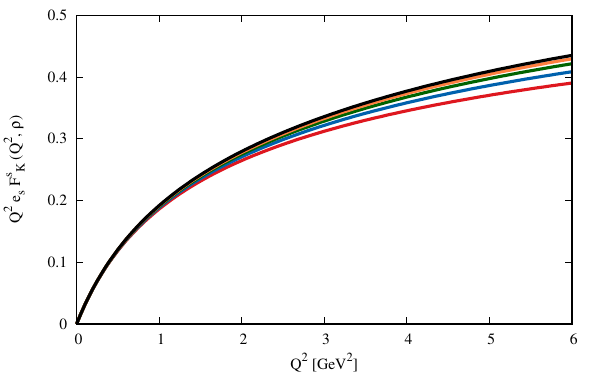}
    \caption{Same as in Fig.~\ref{fig7}, but for $e_s F^s_K (Q^2, \rho_B)$ multiplied by $Q^2$.}
    \label{fig10}
\end{figure}

%======================================================
Results for the total pion and kaon electromagnetic form factors and their quark sector form factors in the free space and nuclear medium are depicted in Figs.~\ref{fig4}-\ref{fig10}. Figure~\ref{fig4} shows the total pion electromagnetic form factors for different nuclear matter densities. The red solid line represents the free space total pion electromagnetic form factor. The result for the total pion electromagnetic form factor in free space is in excellent agreement with the experimental data given in Refs.~\cite{Amendolia:1984nz,JeffersonLab:2008jve} and other theoretical results~\cite{Hutauruk:2016sug}. In Fig.~\ref{fig4}, the pion electromagnetic form factors for nuclear matter densities $\rho/\rho_0 = 0.50, 1.00, 1.50,$ and 2.00, represented by blue, green, orange, and black solid lines, respectively, are shown. It reveals that, as the nuclear matter density increases, the pion electromagnetic form factors decrease, as expected.

To more clearly see the nuclear medium effect in the pion electromagnetic form factors, we compute the pion electromagnetic form factors multiplied by $Q^2$ as shown in Fig.~\ref{fig5}. Also, the result for the up quark sector electromagnetic form factor of the pion multiplied by $e_u Q^2$ is shown in Fig.~\ref{fig6}, where $e_u$ is the elementary electric charge for the up quark. The significant contribution of the up quark sector form factor to the total pion electromagnetic form factor can be also seen in Fig.~\ref{fig6}. Following the total pion electromagnetic form factor, the up quark sector form factor decreases with the increase of the nuclear matter density. It is worth noting that the pion electromagnetic form factors in free space and nuclear medium reproduce $F_\pi (Q^2, \rho) \approx $ 1/Q$^{2}$ for $Q^2 \rightarrow \infty$. At higher $Q^2$, the $Q^2 F_\pi (Q^2,\rho_B=0)$ begins to show plateau with $Q^2F_\pi (Q^2, \rho_B) \simeq$ 0.49 for free space. For $\rho_B/\rho_0 =$ 2.0, $Q^2 F_\pi (Q^2, \rho_B) \simeq$ 0.405.
%%%%%%%%%%%%%%%FIG4%%%%%%%%%%%%%%%%%%
%%%%%%%%%%%%%%%%%%%%%%%%%%%%%%%%%%%%%%%%%%
%%%%%%%%%%%%FIG 5%%%%%%%%%%%%%%%%%%%%%
%%%%%%%%%%%%%%%%%%%%%%%%%%%%%%%%%%%%%%%
%%%%%%%%%%%%%%FIG 6%%%%%%%%%%%%%%%%%%%
%%%%%%%%%%%%%%%%%%%%%%%%%%%%%%%%%%%%%

Besides the pion electromagnetic form factor in free space and nuclear medium, we also compute the kaon electromagnetic form factor in free space and nuclear medium. The results are shown in Fig.~\ref{fig7}. In comparison with the low $Q^2$ existing data ($0 < Q^2 < 0.1$ GeV$^2$)~\cite{Amendolia:1986ui}, the kaon electromagnetic form factor in free space is in excellent agreement with the data~\cite{Amendolia:1986ui}. As in the pion case, we also compute the kaon electromagnetic form factor at $\rho_B/\rho_0 =$ 0.50 (blue solid line), 1.00 (green solid line), 1.50 (orange solid line), and 2.00 (black solid line), respectively. It appears that the kaon electromagnetic form factor decreases as the nuclear matter density increases. The tendency of the kaon electromagnetic form factor is almost similar to the case of the pion. However, they differ in the quark sector form factors. 

For the up quark sector form factors, kaon and pion indicate the same feature, where they increase as the nuclear matter density increases as can be seen in Fig.~\ref{fig9}. However, they have different magnitudes. Similar to the pion case, in Fig.~\ref{fig8}, we also show our results for the kaon electromagnetic form factors multiplied by $Q^2$ for different densities. It also indicates that the kaon electromagnetic form factor at larger $Q^2 \rightarrow \infty$ (asymptotic feature) satisfies $F_K (Q^2, \rho_B) \approx$ 1/Q$^2$ with $Q^2 F_K (Q^2, \rho_B=0) \simeq$ 0.55.

In addition, results for the in-medium up and strange quark sector form factors of the kaon multiplied by $Q^2$ for different densities are shown in Figs.~\ref{fig9} and~\ref{fig10}, respectively. Similar to the up quark sector form factor of the pion in free space and nuclear matter density in Fig.~\ref{fig6}, the up quark sector form factor of the kaon in nuclear medium decreases as the nuclear matter density increases. However, the up quark sector form factor of the kaon decreases faster than that of the pion. Furthermore, we observe the up quark sector form factor for the pion is larger than that of the kaon for the corresponding nuclear matter density as clearly shown in Figs.~\ref{fig6} and~\ref{fig9}, respectively.

Results for the in-medium modifications of the strange sector form factor for the kaon multiplied by $Q^2$ are depicted in Fig.~\ref{fig10}. Interestingly, the results for the strange quark sector form factor differ from the up-quark sector form factor of the kaon. It clearly shows that the strange quark form factors increase as the nuclear matter density increases. The increasing rate is rather slow for different densities. 
%The increasing strange quark form factor weighted by the charges for the kaon compensates for the decreasing up quark sector form factor for the kaon to result in the total kaon electromagnetic form factor satisfying $F_K (Q^2, 
%\rho_B) \approx $ 1/Q$^2$. 
Also, we observe, in free space, that the strange quark sector form factor begins to contribute dominantly at $Q^2 \geq 1.6$ GeV$^2$ to the total electromagnetic form factor of the kaon, which is consistent with that found in Ref.~\cite{Hutauruk:2016sug}. 

It is worth noting that by comparing the in-medium modifications of the kaon electromagnetic form factor calculated by using the hybrid NJL-QMC model~\cite{Hutauruk:2019was} and LF-QMC model~\cite{Yabusaki:2023zin} to the present result of this work, we found that the nuclear medium kaon form factor is not too sensitive to the medium modifications of the kaon mass. In Refs.~\cite{Hutauruk:2019was,Yabusaki:2023zin}, the medium modifications of the kaon masses decrease as the nuclear matter density increases, which is in contrast to our present result, where the medium modifications of the kaon mass increase as the nuclear matter density increases as can be seen in Fig.~\ref{fig1}. The decreasing kaon mass in Refs.~\cite{Hutauruk:2019was,Yabusaki:2023zin} originates from the kaon potential being less repulsive. However, this can be improved by rescaling the repulsive vector potential to some extent (e.g., by rescaling with a factor of $(1.4)^2$) as in Refs.~\cite{Tsushima:1997df,Tsushima:2000re}.

Now, we are in the position to compute the charge radii of the pion and kaon as well as their sector form factors in free space and nuclear medium, which are calculated from the derivative of the form factors with respect to $Q^2$, as formulated in Eq.~(\ref{eq:radiiNJL}).
%%%%%%%%%%%%%%%%FIG 7%%%%%%%%%%%%%%%%%%
%%%%%%%%%%%%%%%%%%%%%%%%%%%%%%%%%%%%%%%%%%

%%%%%%%%%%FIG 8 %%%%%%%%%%%%%%%%%%%%
%%%%%%%%%%%%%%%%%%%%%%%%%%%%%%%%%%%%%

%%%%%%%%%%%FIG 9%%%%%%%%%%%%%%%%%%%
%%%%%%%%%%%%%%%%%%%%%%%%%%%%%%%%%%

%%%%%%%%%%%FIG 10%%%%%%%%%%%%%%%%
%%%%%%%%%%%%%%%%%%%%%%%%%%%%%%%

%======================================================
\subsubsection{Charge radius}
%======================================================
Using the expression for the in-medium modifications of the kaon and pion electromagnetic form factors in Eq.~(\ref{eq:NJL15}), respectively, the root-mean-square charge radius of the pion and kaon for various densities can be obtained from
%%%%%%%%%
\begin{eqnarray}
    \langle r_{\pi(K)}^2 \rangle &=& - 6 \frac{dF_{\pi (K)}(Q^2,\rho_B)}{dQ^2} \Bigg|_{Q^2 =0}. 
    \label{eq:radiiNJL}
\end{eqnarray}

%%%
\begin{table}[t]
	\begin{ruledtabular}
		\renewcommand{\arraystretch}{1.1}
		\caption{Kaon and pion charge radii in free space and nuclear medium and their quark sector form factors. All radii are in units of fm.}
		\label{tab:NJL3}
		\begin{tabular}{cccccc}
   $\pi^+$  & $\rho_B/\rho_0$  & $r_\pi$  & $r_u$  & $r_d$  & $r_{\rm{expt.}}$~\cite{ParticleDataGroup:2016lqr} \rule[-2ex]{0pt}{4ex}\\  \hline
		 &  0.0 & 0.629 & 0.629  & $-0.629$ & $0.672 \pm 0.008$\\ 
          &  0.5 & 0.664 & 0.664  & $-0.664$ \\ 
            &  1.0 & 0.694 & 0.694  & $-0.694$ \\ 
              &  1.5 & 0.714 & 0.714  & $-0.714$ \\ 
          &  2.0 & 0.730 & 0.730  & $-0.730$ \\ \hline \hline
          \\[-2ex]
          $K^+$ &  $\rho_B/\rho_0$ & $r_K$  &  $r_u$  &  $r_s$ & $r_{\rm{expt.}}$~\cite{ParticleDataGroup:2022pth} \rule[-2ex]{0pt}{4ex} \\ \hline 
           &  0.0 & 0.586 & 0.646  & $-0.441$ & $0.560 \pm 0.031$\\ 
          &  0.5 & 0.614 & 0.686  & $-0.434$ \\ 
            &  1.0 & 0.638& 0.719 & $-0.434$ \\ 
              &  1.5 & 0.655 & 0.742  & $-0.433$ \\  
            &  2.0 & 0.668 & 0.759  & $-0.432$ \\  \hline \hline
            \\[-2ex]
          $K^0$  &  $\rho_B/\rho_0$ &  $r_K$ &  $r_d$  & $r_s$ & $r_{\rm{expt.}}$~\cite{ParticleDataGroup:2018ovx} \rule[-2ex]{0pt}{4ex}\\ \hline 
            &  0.0 & $-0.272$ & 0.646  & $-0.441$ & $-0.277 \pm 0.018$ \\ 
          &  0.5 & $-0.307$ & 0.686  & $-0.434$ \\ 
            &  1.0 & $-0.331$ & 0.719  & $-0.434$ \\ 
              &  1.5 & $-0.348$ & 0.742  & $-0.433$ \\  
            &  2.0 & $-0.361$ & 0.759  & $-0.432$ 
		\end{tabular}
		\renewcommand{\arraystretch}{1}
	\end{ruledtabular}
\end{table}
%%%

\begin{figure}[t]
    \centering
    \includegraphics[width=1\linewidth]{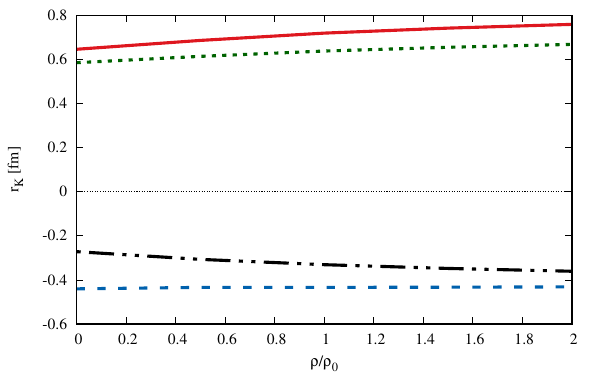}
    \caption{Kaon radius and its quark sector charge radius contributions in the nuclear medium. The solid, dotted, dashed, and dash-dotted lines represent respectively $r_u$, $r_{K^+}$, $r_{s}$, and $r_{K^{0}}$.}
    \label{fig11}
\end{figure}

\begin{figure}[t]
    \centering
    \includegraphics[width=1\linewidth]{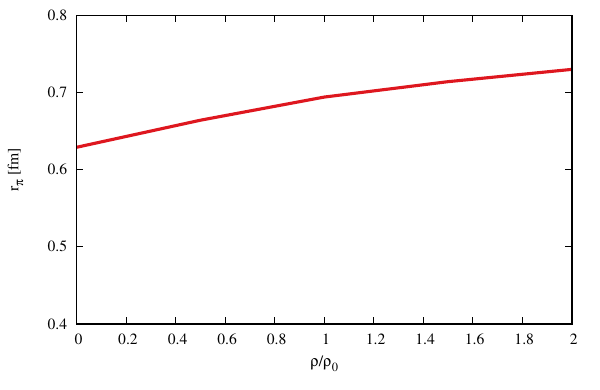}
    \caption{Pion charge radius in free space and the nuclear matter.}
    \label{fig12}
\end{figure}

Results for the kaon and pion charge radii in free space and nuclear medium are shown in Figs.~\ref{fig11} and~\ref{fig12}. Figure~\ref{fig11} shows the kaon and its quark sector charge radii. In free space, the kaon charge radius is in good agreement with the experimental data~\cite{ParticleDataGroup:2022pth}. The kaon charge radius (solid line) in a nuclear medium increases as the nuclear matter density increases. Our results are also consistent with other calculations obtained by using different approaches~\cite{Hutauruk:2019was}.

In Fig.~\ref{fig11}, it is shown that the contribution of the up-quark charge radius (dotted line) increases as the nuclear matter density increases. The numerical values of the up-quark charge radius contributions can be also seen in Table.~\ref{tab:NJL3}. Also, we show the strange-quark charge radius (dashed line) contribution to the total kaon charge radius. It reveals that the strange quark charge radius contribution does not significantly vary in the nuclear medium, as expected. This is because the strange quark mass does not significantly change in the nuclear medium and the strange quark weakly interacts with vector and scalar mean fields in the nuclear medium. 

Besides the positively charged kaon radius result, we also provide the result for the charge radius for the neutral kaon (dash-dotted line) in free space and nuclear medium. It shows that the charge radius for the neutral kaon decreases as the nuclear matter density increases as shown in Fig.~\ref{fig11}. 
For the positively charged and neutral kaons, the contributions of strange and up quarks charge radii are similar, as indicated in Table.~\ref{tab:NJL3}. Only their total charges are different.

Finally, we show our result for the pion charge radius in free space and nuclear medium in Fig.\ref{fig12}. The free space charge radius for the pion reproduces the experimental data~\cite{ParticleDataGroup:2016lqr} nicely, while the nuclear medium charge radius increases as the nuclear matter density increases. This result is consistent with other calculations available in the literature~\cite{Hutauruk:2018qku,deMelo:2014gea}.

%======================================================
\section{Summary} \label{sec:summary}
%======================================================
In the present work, we have investigated the in-medium modifications of the kaon and pion properties and structures as well as their charge radii in the framework of the NJL model with the help of the Schwinger proper-time regularization scheme to simulate the effect of QCD confinement. The nuclear medium effect is also computed with the same NJL model consistently, where the chiral symmetry-breaking partial restoration is captured in the model \textit{via} the chiral quark condensate as an order parameter. In this work, we compute the free space pion and kaon properties and structures as well as the nuclear medium. We then compare our free space results with the free space experimental data and other theoretical calculations. We found that the free space kaon and pion electromagnetic form factors are in good agreement with the experimental data, which is followed by the kaon and pion charge radii, which result in good agreement with data. 

With these good free space results, we predict the kaon and pion electromagnetic form factors in the nuclear medium. It is found that the kaon electromagnetic form factor decreases as the nuclear matter density increases. Similarly, the pion electromagnetic form factor in the nuclear medium also decreases as the nuclear matter density increases. Their quark sector form factor contributions for the pion and kaon are also investigated.

Also, we predict the kaon and pion charge radii in the nuclear medium. It is found that the kaon charge radius in the nuclear medium increases as the nuclear matter density increases. A similar trend is also shown by the pion charge radius.

Another interesting result of this work is the in-medium modifications of the axial-vector $g_A^*$ coupling that is calculated via GMOR. We found that $g_A^*$ does not significantly change in the nuclear medium, where at $\rho_B=\rho_0$, the reduction of $g_A$ is estimated about 0.2\% relative to
that in free space, indicating a very small quenching factor.

The present studies of the pion and kaon properties in free space and nuclear medium are very useful to provide valuable and relevant information to the deeply bound pionic and kaonic atom experiments that are planned to be measured at J-PARC in Japan~\cite{Aoki:2021cqa}. Also, the results of this study may provide useful and helpful guidance for the lattice QCD. Next, for future work, we plan to apply this consistent NJL model in calculating the electromagnetic form factors of spin-1 vector mesons with equal and unequal quark masses in the nuclear medium. Related works are still in progress and will appear elsewhere.

~

%======================================================
\section*{ACKNOWLEDGEMENTS}
%======================================================
 This work was supported by the National Research Foundation of Korea (NRF) grants funded by the Korean government (MSIT) Nos. 2018R1A5A1025563, 2022R1A2C1003964, and 2022K2A9A1A0609176, and by the PUTI Q1 Grant from the University of Indonesia under contract No. NKB-442/UN2.RST/HKP.05.00/2024.
 %=====================================================

%======================================================
%\newpage 

%======================================================
\end{document}